\documentclass[prl,twocolumn,aps,amsfonts,showpacs,longbibliography,notitlepage,nofootinbib,superscriptaddress]{revtex4-1}

\usepackage{graphicx}
\usepackage[dvipsnames]{xcolor}
\usepackage{rotating}
\usepackage{amsmath,amssymb,graphics,amsthm}
\usepackage{amsfonts,dsfont,mathtools}
\usepackage{bbm}
\usepackage{bm}
\usepackage{array}
\usepackage{appendix}
\newcolumntype{P}[1]{>{\centering\arraybackslash}p{#1}}
\usepackage{multirow}
\usepackage{physics}
\usepackage{tikz-cd}
\usepackage{adjustbox}
\usepackage{stmaryrd}
\usepackage{tikz}
\usetikzlibrary{shadows.blur}

\usepackage[colorlinks=true, urlcolor=violet, linkcolor=blue, citecolor=red, hyperindex=true, linktocpage=true]{hyperref}
\usepackage[capitalise,compress]{cleveref}

\allowdisplaybreaks

\newtheorem{thm}{Theorem}
\makeatletter
\renewcommand{\p@subsection}{}
\renewcommand{\p@subsubsection}{}
\makeatother

\newcommand\bea{\begin{eqnarray}}
\newcommand\eea{\end{eqnarray}}
\newcommand\be{\begin{equation}}
\newcommand\ee{\end{equation}}
\newcommand\bes{\begin{subequations}}
\newcommand\ees{\end{subequations}}
\newcommand\bed{\begin{displaymath}}
\newcommand\eed{\end{displaymath}}
\newcommand\beal{\begin{aligned}}
\newcommand\eeal{\end{aligned}}
\newcommand\bew{\begin{widetext}}
\newcommand\eew{\end{widetext}}
\newcommand\beit{\begin{itemize}}
\newcommand\eeit{\end{itemize}}
\def\bea{\begin{array}}
\def\eea{\end{array}}
\newcommand\been{\begin{enumerate}}
\newcommand\eeen{\end{enumerate}}

\def\i{\text{i}}
\newcommand{\id}[0]{\mathds{1}}

\usepackage{verbatim}

\usepackage{tikz}
\usetikzlibrary{quantikz2}

\definecolor{red1}{rgb}{0.76, 0.23, 0.13}
\definecolor{green1}{rgb}{0.0, 0.5, 0.0}
\usepackage{pifont} % http://ctan.org/pkg/pifont

\usepackage{booktabs}

\usepackage[T1]{fontenc}
\usepackage[dvipsnames]{xcolor}
\usepackage{multirow}
\usepackage{booktabs} 
\usepackage{caption}
\usepackage[normalem]{ulem}

\begin{document}

\title{Robustness-Runtime Tradeoff for Quantum State Transfer}

\author{Twesh Upadhyaya}
\email{tweshu@umd.edu}
\affiliation{Joint Center for Quantum Information and Computer Science, University of Maryland and NIST, College Park, MD 20742, USA}
\affiliation{Department of Physics, University of Maryland, College Park, MD 20742, USA}
\affiliation{Joint Quantum Institute, University of Maryland and NIST, College Park, MD 20742, USA}

\author{Yifan Hong}
\affiliation{Joint Center for Quantum Information and Computer Science, University of Maryland and NIST, College Park, MD 20742, USA}
\affiliation{Joint Quantum Institute, University of Maryland and NIST, College Park, MD 20742, USA}

\author{T. C. Mooney}
\affiliation{Joint Center for Quantum Information and Computer Science, University of Maryland and NIST, College Park, MD 20742, USA}
\affiliation{Joint Quantum Institute, University of Maryland and NIST, College Park, MD 20742, USA}

\author{Alexey V. Gorshkov}
\affiliation{Joint Center for Quantum Information and Computer Science, University of Maryland and NIST, College Park, MD 20742, USA}
\affiliation{Joint Quantum Institute, University of Maryland and NIST, College Park, MD 20742, USA}

\date{February 25, 2026}

\begin{abstract}
    Quantum state transfer is the primitive of transporting an unknown state on one site of a lattice to another. 
    Using power-law interactions, recent state transfer protocols achieve speedup by utilizing the intermediate ancilla sites. However, these protocols require the ancillas to be in a perfectly initialized state, which, due to noise or imperfect control, may not be the case.
    In this work we introduce the \emph{robustness} of a state transfer protocol, which quantifies the protocol's tolerance to error in the initial ancilla state. 
    In the Heisenberg picture, state transfer grows operators supported on the final site such that they no longer commute with all operators on the starting site. 
    We prove that this robustness tightly bounds the Schatten $p$-norms of these commutators between initial and final-site operators. 
    This generalizes the known cases of $p=\infty$ and $p=2$, which govern completely state-dependent and state-independent state transfer respectively, demonstrating that intermediate values of $p$ govern partially state-dependent state transfer.
    In conjunction with existing power-law light cones, our result gives new minimum runtimes for partially state-dependent protocols which, in certain regimes, are parametrically better than existing bounds.
    We introduce new robust state transfer protocols, charting the landscape between complete state-dependence and state-independence. 
\end{abstract}

\maketitle

\begin{figure}[t]
    \centering
\begin{tikzpicture}[scale=0.7]

\def\a{1.6}        
\def\Rblue{0.48}

\definecolor{softblue}{RGB}{110,155,205}
\definecolor{softred}{RGB}{215,115,115}
\definecolor{softgreen}{RGB}{120,190,145} 

\tikzset{
    bluequbit/.style={
        fill=softblue,
        opacity=0.85
    }
}

\newcommand{\psiQubit}[2]{

    \path[fill=softgreen, opacity=0.85] (#1,#2) circle (\Rblue);

    \foreach \k in {1,...,12} {
        \pgfmathsetmacro{\rk}{\Rblue * \k / 12}
        \pgfmathsetmacro{\ok}{0.75 * exp(-0.25*\k)}
        \path[fill=white, opacity=\ok] (#1,#2) circle (\rk);
    }

    \node at (#1,#2) {$\ket{\psi}$};
}

\psiQubit{0}{0}

\path[bluequbit] (\a,\a) circle (\Rblue);
\path[bluequbit] (2*\a,0) circle (\Rblue);

\foreach \x/\y in {\a/0, 0/\a, 0/2*\a} {
    \foreach \k in {1,...,12} {
        \pgfmathsetmacro{\rk}{\Rblue * \k / 12}
        \pgfmathsetmacro{\ok}{0.8 * exp(-0.15*\k)}
        \path[fill=softred, opacity=\ok]
            (\x,\y) circle (\rk);
    }
}

\node[opacity=0.55,scale=0.8] at (2.8*\a,0) {\Huge $\cdots$};

\node[opacity=0.55, rotate=90,scale=0.8] at (0.037,2.8*\a) {\Huge $\cdots$};

\node[opacity=0.55, rotate=45,scale=0.8] at (1.7*\a,1.7*\a) {\Huge $\cdots$};

\path[bluequbit] (3.2*\a,3.2*\a) circle (\Rblue);

\coordinate (qBL) at (0,0);
\coordinate (qTR) at (3.2*\a,3.2*\a);

\coordinate (ctrlA) at (1.1*\a, 2.85*\a);
\coordinate (ctrlB) at (2.65*\a, 3.2*\a);

\tikzset{transferarrow/.style={
    -{Triangle[length=8pt,width=9pt]},
    line cap=round, line join=round
}}

\draw[transferarrow,
      line width=3pt, black!20, opacity=1,
      shorten <=\Rblue cm, shorten >=\Rblue cm]
  ([xshift=2.6pt,yshift=-2.6pt]qBL)
    .. controls ([xshift=2.6pt,yshift=-2.6pt]ctrlA)
    and       ([xshift=2.6pt,yshift=-2.6pt]ctrlB)
    .. ([xshift=2.6pt,yshift=-2.6pt]qTR);

\draw[transferarrow, draw=softgreen!80!black,
      line width=2.4pt, opacity=1,
      shorten <=\Rblue cm, shorten >=\Rblue cm]
  (qBL) .. controls (ctrlA) and (ctrlB) .. (qTR);

\end{tikzpicture}
\caption{Quantum state transfer is the task of transporting an unknown state from one lattice site to another. The states of intermediate sites may not be fully specified.}
    \label{fig:keyimage}
\end{figure}
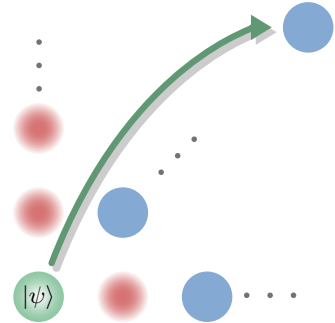

\emph{Quantum state transfer} is a fundamental task in quantum information processing \cite{Bose_2003,Christandl2004Perfect,Christandl2005Perfect,Yung2005Perfect,Franco2008Perfect}: given a quantum state on one site, transfer it to another site using available interactions (Figure~\ref{fig:keyimage}). It is widely used as a subroutine to generate the long-range entanglement needed in many applications of quantum information science. In distributed quantum computing \cite{Cacciapuoti_2020, Azuma_2023, Caleffi_2024} for instance, state transfer is used to entangle distant nodes in a network \cite{Briegel_1998, Gottesman_2012, huhtanen_2026}. In quantum error correction, state transfer can realize the long-range interactions required to implement exotic error-correcting codes \cite{Berthusen_2025_2D}. In a fault-tolerant quantum processor, state transfer can be used as a logical subroutine to move logical information between memory and computational code blocks \cite{Xu_2024_constant, viszlai2024, cross2025improved, he2025extractors}. 

There are several natural questions to explore when analyzing state transfer, and the answers all depend sensitively on the setup and available gadgets at hand. One of the most essential questions is how fast state transfer can be achieved over a certain distance. 
In the Heisenberg picture, operators on the final site spread to have nontrivial support on the initial site. Commutation relations between this time-evolved final operator and operators on the initial site thus provide constraints on state transfer \cite{Epstein2017}.
It is well-known that the asymptotic speed of unitary state transfer is governed by Lieb-Robinson bounds \cite{Lieb_1972, Bravyi_2006_LRB, Nachtergaele_2006, hastings2010locality, Nachtergaele_2010, Kliesch_2014, Chessa_2019, Wang_2020, Chen_2021, Chen2023}, which roughly capture the dynamical spread of local operators under time evolution. In their original titular paper \cite{Lieb_1972}, Lieb and Robinson showed the existence of a linear light cone for dynamics under finite-range and exponentially decaying interactions. As a consequence, under this setup, there is no state transfer protocol that is parametrically faster than locally hopping (swapping) the desired state from the initial to the final site.

However, under \emph{power-law} dynamics \cite{Monroe2021Programmable, Saffman2010Quantum, Doherty2013Nitrogen, Gadway2016Strongly}, one can generate large entangled states on the intermediate sites and leverage their long-range collective enhancements to achieve state transfer faster than with nearest-neighbor hopping \cite{Guo_2020, Tran2021a, Hong_2021_fast, Yin_2025_all}.
During the culmination of breakthroughs for power-law Lieb-Robinson bounds \cite{Hastings_2006, FossFeig_2015, Matsuta_2016, Eldredge_2017, Tran_2019, Chen_2019, Luitz_2019, Else_2020, Chen_2019_1D, Kuwahara_2020, Tran_2020_hierarchy, Tran2021lrb, Gong_2023} and state transfer protocols \cite{Guo_2020, Tran2021a, Hong_2021_fast}, it was discovered that light-cone shapes depend on how operator spreading is quantified. In particular, traditional Lieb-Robinson bounds examined commutators using the operator norm, which measures the largest singular value. Using a different norm that takes into account all the singular values, such as the Frobenius norm, results in a qualitatively different light cone \cite{Tran_2020_hierarchy, Chen2021}. But if the operator norm governs unitary state transfer, then what are the physical interpretations of the other norms? In the same paper \cite{Tran_2020_hierarchy}, it was shown that the Frobenius norm lower bounds the state transfer time when all intermediate sites are initialized in unknown states. This separation between different norms is a special feature of long-range interactions and is absent in local interactions. Indeed, the optimal nearest-neighbor hopping protocol for local interactions is agnostic to the initial state on the intermediate sites. However, the collective enhancement in the long-range protocols will be ruined if the intermediate entangled states are imperfectly prepared.

In this paper, we establish a precise connection between the \emph{robustness} of state transfer and the growth of commutators with respect to a family of norms. By robustness, we mean roughly the largest fraction of uninitialized intermediate sites that the protocol can tolerate.
We prove that the larger this fraction is, the larger the commutators need to grow under the family of Schatten $p$-norms. Loosely speaking, on a $d$-dimensional lattice with $L$ sites, a state-transfer protocol from site $i$ to $f$ that is agnostic to the state of $k$ intermediate sites obeys a commutator lower bound of
\begin{align}
    \norm{[X_f(t), Z_i]}_p \gtrsim 2^{(k-L)/p} \, .
\end{align}
See Theorem \ref{thmmainpnorm} for the formal statement. We show that this lower bound is tight by constructing a saturating state transfer protocol. 
These statements are made without any reference to locality. We then combine our bounds with existing $p$-norm light cones \cite{Chen2021} to derive new bounds on robust, power-law state transfer.  
Finally, we introduce and analyze various robust state transfer protocols and compare them to our bounds.
The notion of robust state transfer is especially relevant in solid-state quantum computing architectures, where intermediate physical qubits may be imperfectly initialized~\cite{Yao2011, Tsomokos_2007}.

\emph{State transfer.}---Consider a $d$-dimensional lattice of $L$ $D$-level qudits in with a distance measure $\mathcal{D}$ between sites. The Hilbert space is denoted by $\mathcal{H}_L = (\mathbb{C}^D)^{\otimes L}$. One site is designated the initial site $i$ and the rest \emph{ancilla} sites $\text{anc}$. One of the ancilla sites, typically one furthest from $i$, is designated as the final target site $f$. We use $\text{anc}'$ to refer to all sites except $f$ (this includes $i$). Where it is clear from context, we omit the subscripts labelling the different sets of qudits.

We define an \emph{$\mathcal{S}$-robust state transfer protocol} to be a Hamiltonian that transfers an arbitrary state from the initial site to the final site whenever the (pure) ancilla state is in the subspace $\mathcal{S} \subset \mathcal{H}_{L-1}$. That is, the unitary $U$ generated by the protocol satisfies
\begin{equation}
    U \left(\ket{\psi}_i \otimes \ket{\Phi}_{\rm anc}\right) = \ket{\Phi'}_{\rm anc'} \otimes \ket{\psi}_f
\end{equation}
for every $\ket{\psi}\in \mathbb{C}^D$ and every $\ket{\Phi} \in \mathcal{S}$. Linearity immediately implies that the state transfer protocol works for any mixed state with support only on $\mathcal{S}$. Denote by $\Pi_{\mathcal{S}}$ the projector onto the subspace. State-independent state transfer corresponds to $\Pi_{\mathcal{S}}=\id$, while the completely state-dependent state transfer protocol in Ref.~\cite{Tran2021a} corresponds to $\Pi_{\mathcal{S}}=\dyad{\bar{0}}$, where we use the notation $\ket{\bar{0}}$ to represent the product state $\ket{00...0}$. 
Our main results hold for general geometries and local Hilbert space dimension $D$. For simplicity we restrict to qubits ($D=2$) in the following sections but present the extension to qudits in the Supplemental Material (SM)~\cite{suppmat}.

We have defined state transfer in the Schrödinger picture, but it is also useful to consider state transfer in the Heisenberg picture. We denote by $X,Y,Z$ the qubit Pauli matrices. From Ref.~\cite{Hong2024}, an equivalent condition for $U$ to be a $\mathcal{S}$-robust state transfer protocol is 
\begin{align}\label{stabilizer1}
U^\dagger X_f U &=X_i S_x \quad \text{and} \\
U^\dagger Z_f U &=Z_i S_z,\label{stabilizer2}
\end{align}
where $S_{x,z}$ are ``stabilizer''\footnote{Note that this definition of ``stabilizer'' is more general than the one usually associated with Clifford simulation. Namely, we allow $S_x$ and $S_z$ to be non-Paulis.} unitaries that obey 
\begin{equation}\label{subspace}
    S_x (\ket{\psi}_i \otimes \ket{\Phi}_{\rm anc})  = S_z (\ket{\psi}_i \otimes \ket{\Phi}_{\rm anc}) = \ket{\psi}_i \otimes \ket{\Phi}_{\rm anc} 
\end{equation} for all $\ket{\Phi}\in\mathcal{S}$ and $\ket{\psi}\in \mathbb{C}^D$.
(The transfer of more than one qubit is encompassed by adding an analogous condition for each additional transferred qubit, see SM for details.) For example, state-independent state transfer is achieved only when $S_x=S_z=\id$. Note that by Hermiticity,
\begin{equation}\label{herm}
X_i S_x = S_x^\dagger X_i,    
\end{equation}
and similarly for $S_z$.
Also note that the stabilizers do not necessarily have trivial support on the initial site. For example, a stabilizer of the form $S_x=\id \otimes \dyad{\bar{0}}+ Z\otimes (\id-\dyad{\bar{0}})$ is compatible with a state-dependent protocol. For qudits, substituting $X$ and $Z$ with generalized Pauli matrices in \eqref{stabilizer1} and \eqref{stabilizer2}, with the stabilizers still obeying \eqref{subspace}, gives necessary and sufficient conditions for state transfer (see SM).

\emph{Norms.---}The Schatten $p$-norm of a Hermitian or anti-Hermitian operator is defined as the $p$-norm of the absolute values of the eigenvalues \cite{bhatia_2013}. In keeping with previous literature, we use $p$-norms renormalized by the dimension of the Hilbert space. For an operator $\hat{O}$ with eigenvalues $\lambda_i$ acting on a Hilbert space $\mathcal{H}$, in this convention, the $p$-norm is thus defined as
\begin{equation}
\big\Vert\hat{O}\big\Vert_p\coloneqq \abs{\mathcal{H}}^{-1/p} \left(\sum \abs{\lambda_i}^p\right)^{\frac{1}{p}}.    
\end{equation}
The limit $p\rightarrow \infty$ gives the operator norm, while $p=2$ gives the Frobenius norm. The renormalization ensures that $\norm{\id}_p = 1$.
Due to the renormalization, the usual ordering between $p$-norms is inverted, $\big\Vert\hat{O}\big\Vert_q \leq \big\Vert\hat{O}\big\Vert_p$ for $q$ \emph{less} than $p$.

\emph{Power-Law Light Cones.---}
Lieb-Robinson bounds constrain the growth of operators under local Hamiltonians. 
Specifically, they bound the operator norm of the commutator between two operators $A_i$ and $U^\dagger B_f U$ in terms of evolution time and distance $\mathcal{D}(i,f)$.
Historically, only the operator norm was used for such bounds, as opposed to light cones bounding the $p$-norm of the commutator---one of the reasons being that for, say, nearest-neighbour Hamiltonians, these different light cones have the same scaling. However, for \emph{power-law} Hamiltonians (see below) the light cones are in fact different \cite{Tran_2020_hierarchy}. Tight light cones are known for $p=\infty$ for any $d$ and for $p=2$ for $d=1$, which we summarize in the End Matter.
Power-law Hamiltonians model important classes of interactions in quantum information \cite{Monroe2021Programmable, Saffman2010Quantum, Doherty2013Nitrogen, Gadway2016Strongly}, and are of the form 
\begin{equation}
    H= \sum_i h_i(t) + \sum_{i<j} h_{ij}(t),
\end{equation}
where the local onsite terms $h_i(t)$ are arbitrary and the interaction terms obey $\norm{h_{ij} (t)}_\infty \leq \frac{1}{\mathcal{D}(i,j)^\alpha}$ for some $\alpha >0$. $U=\mathcal{T} e^{-i \int_0^t H(\tau)d\tau} $ is the unitary representing time evolution under $H$ for time $t$.

A general light cone for power-law Hamiltonians would be of the form 
\begin{equation}\label{generallightcone}
    \norm{[U^\dagger B_f U, A_i]  }_p >\delta \implies t \gtrsim g(\alpha,r,p,\delta),
\end{equation}
where the function $g$ is decreasing in $\alpha$ (as the interaction strength increases, operations can be performed faster), increasing in $r$ (farther distances take longer), decreasing in $p$ (monotonicity of $p$-norms), and increasing in $\delta$ (bigger norms take longer to achieve). 
In 1D, a light cone has been derived for general $p$, though it is not tight. From Eq.(4.18) of \cite{Chen2021}, with the result rewritten for renormalized $p$-norms and a minor typo corrected, we have that for $p\geq2$
\begin{equation}\label{pnormlightcone}
    \norm{[U^\dagger X_f U,Z_i]  }_p>\delta  \implies t \geq \delta  \frac{R(r)}{ \sqrt{p} C}, 
\end{equation}
with
\begin{equation}\label{eq:R(r)}
    R(r)=\begin{cases}
    r^{\alpha-3/2} &\quad 3/2<\alpha < 5/2\\
    r/\ln^{3/2}(r) &\quad \alpha = 5/2\\
    r &\quad \alpha>5/2    .
    \end{cases}
\end{equation}

\emph{State transfer with noisy ancillas.}---We now state our main results connecting the robustness of state transfer to the growth of commutator norms. Commutators between an operator on the initial site and a time-evolved operator on the final site provide necessary conditions on state transfer. Taking the commutator with $Z_i$ on both sides of Eq.~\eqref{stabilizer1} gives 
\begin{equation}\label{maincommutator}
    [U^\dagger X_f U,Z_i] =[X_i S_x,Z_i].
\end{equation}
(A similar equation can be written for $S_z$.) The norm of the left-hand side can be bounded by light cones. Our main result connects the right-hand side to state transfer robustness:
\begin{thm}\label{thmmainpnorm}
    Any $\mathcal{S}$-robust state transfer protocol satisfies $\norm{[X_i S_x, Z_i]}_p \geq 2\cdot 2^{(-L+1)/p} \abs{\mathcal{S}}^{1/p} $.
\end{thm}
This theorem is a formal statement of the intuition we alluded to earlier: if a state transfer protocol works for more ancilla states, it has to grow a ``bigger'' commutator, but the necessary growth is suppressed by $p$. The proof can be found in the SM.

\begin{figure}
    \centering
    \includegraphics[width=0.45\textwidth]{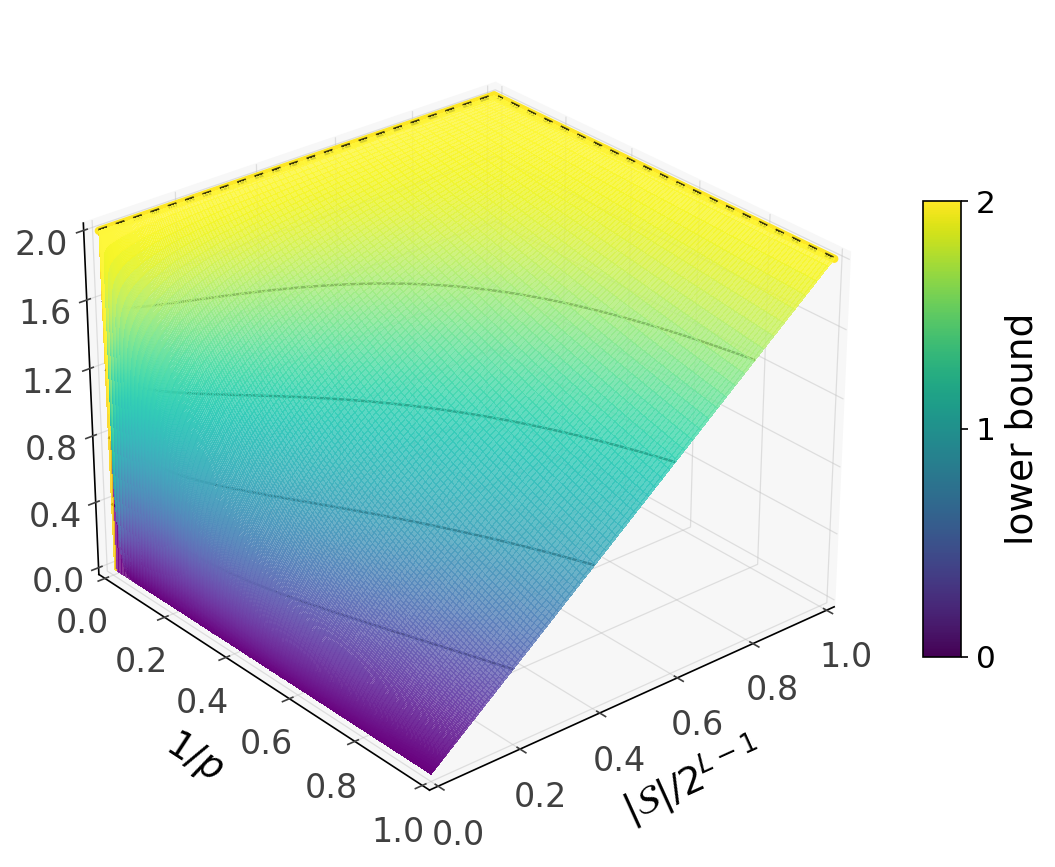}
    \caption{Lower bound from Theorem~\ref{thmmainpnorm}: minimum commutator $p$-norm of any $\mathcal{S}$-robust state transfer protocol, plotted as a function of $\abs{\mathcal{S}}$ and $p$. Maximal values at state-independence ($\abs{\mathcal{S}}=2^{L-1}$) and operator norm ($p=\infty$) are indicated by the dashed lines. }
    \label{fig:surface}
\end{figure}

In Figure~\ref{fig:surface}, we plot the lower bound as a function of $p$ and the Hilbert space fraction $2^{1-L}\abs{\mathcal{S}}$.
Some specific values merit comment. As long as $\abs{\mathcal{S}}\geq1$, in the limit $p\rightarrow\infty$ we get $\norm{[X_i S_x, Z_i]}_p \rightarrow 2 $ as the lower bound from Theorem~\ref{thmmainpnorm} and the upper bound from the Cauchy-Schwarz inequality coincide. This upper limit recovers the known result for the operator norm.  Similarly, when the ancillary dimension is maximal, i.e. $\abs{\mathcal{S}}=2^{L-1}$, all norms are maximal $\norm{[X_i S_x, Z_i]}_p =2 $. This lower limit recovers the known result for state-independent state transfer. 
For $p=1$, the bound grows linearly with the subspace dimension $\norm{[X_i S_x, Z_i]}_1 \geq 2\cdot 2^{-L+1} \abs{\mathcal{S}}$. As $\abs{\mathcal{S}}/2^{L-1}$ goes to zero, the bound goes to zero except at $1/p=0$, where it stays at $2$.

We now show that Theorem~\ref{thmmainpnorm} is in fact tight. For every subspace $\mathcal{S}$, we can construct a state transfer protocol that saturates the bound in Theorem~\ref{thmmainpnorm}, even though the bound only depends on $\abs{\mathcal{S}}$. To be precise, we construct a state transfer unitary---which can be implemented by different Hamiltonians depending on the specific scenario under consideration (e.g. power-law interactions, nearest-neighbor, etc.).  
As a minor technical convenience, we require the subspace projector to be of the form $\Pi_\mathcal{S}=\Pi_\mathcal{S'} \otimes \id_f$, so that $\abs{\mathcal{S}}=2\abs{\mathcal{S'}}$.
This restriction is mild in the sense that it can be satisfied by adding a single uninitialized site. Let SWAP denote the unitary which exchanges two systems, SWAP$ \ket{\psi_1} \ket{\psi_2}=\ket{\psi_2} \ket{\psi_1}$. Let $\tilde{H}$ denote the qubit Hadamard unitary which satisfies $\tilde{H}Z\tilde{H}=X$. Finally, let $\bar{\Pi}_\mathcal{S'}=\id-\Pi_\mathcal{S'}$. Consider the state transfer unitary
\begin{equation}\label{eq:saturating U}
    U= \mathrm{SWAP}_{i,f} \otimes \Pi_\mathcal{S'}  +  (\mathrm{SWAP}_{i,f} \tilde{H}_i) \otimes \bar{\Pi}_\mathcal{S'}\, . 
\end{equation}
Conditioned on the ancillas being in $\mathcal{S'}$, this unitary achieves perfect state transfer. Otherwise, it achieves state transfer up to a Hadamard rotation. This is the worst error from the perspective of commutator-norm growth, since it rotates $Z$, which is maximally noncommuting with $X$, to become $X$.
Explicit computation shows that $S_x$ and $S_z$ share a stabilizer eigenspace of dimension $2\abs{\mathcal{S}'}$ and the resultant commutator norm saturates the bound from Theorem~\ref{thmmainpnorm}. We give details in the SM. 
This state transfer protocol \eqref{eq:saturating U} is frugal in that it grows the commutator as little as possible while achieving robust state transfer.

\emph{Application to systems with local interactions.---}
We have established a general theorem linking the robustness of state transfer to commutator growth. 
This result has not invoked locality---we have only been concerned with the final unitary that is implemented, not how it is actually compiled from a local Hamiltonian. We now take into account locality and the (parametric) runtime of state transfer protocols. Since improved light cones for $p<\infty$ are only known in 1D, in this section we specialize to a chain of $L$ qubits where the initial and final sites are at the opposite ends. If stronger light cones [Eq.~\eqref{generallightcone}] in higher dimensions are discovered, they can immediately be applied in conjunction with our Theorem~\ref{thmmainpnorm} to lower bound state transfer times.

By Theorem~\ref{thmmainpnorm}, a $\mathcal{S}$-robust state transfer protocol with $\abs{\mathcal{S}}=2^k$ must achieve $\delta> 2\cdot 2^{(k-L+1)/p}  $. Thus, the light cone for any $p$ provides a lower bound on the time needed for state transfer. 
Given a light cone of the form in Eq.~\eqref{generallightcone}, the strongest such constraint is
\begin{equation}\label{popt}
    t\gtrsim\max_p g(\alpha,L,p, 2\cdot 2^{(k-L+1)/p} ).
\end{equation}
There is a tradeoff as increasing $p$ contributes to decreasing $g$ by the third argument but increasing $g$ by the fourth argument. 
Although it is not tight, using the light cone in Eq.~\eqref{pnormlightcone}, we can compute the first nontrivial bounds for partially state-dependent state transfer; prior to this work the best bounds were from the operator norm. To do so, we explicitly solve the optimization in Eq.~\eqref{popt} for the specific bound in Eq.~\eqref{pnormlightcone}. 
\begin{thm}\label{thmoptimalp}
\begin{equation}\label{eq:p_star}
    p_* \equiv \arg \max_p \ 2 R(L) \frac{2^{(k-L+1)/p}}{ \sqrt{p} C} = \max\{2,(2 \ln 2) (L-k-1)\}.
\end{equation}    
\end{thm}
We relegate the formal proof to the SM. As expected, for a larger $k$ the optimal $p$ is smaller. 
For $k=L-1$, $p_*=2$, in agreement with the fact that the Frobenius norm gives a largely tight lightcone for state-independent state transfer. For $k=0$, $p_*\rightarrow \infty$ as $L\rightarrow \infty$, in agreement with the fact that the operator norm gives an asymptotically tight lightcone for completely state-dependent state transfer.
Plugging in the optimal value $p_*$ gives us the best bound. Set $C'=2\ln 2$ so that $p_*= C' (L-k-1)$. Assume also from hereon that $k\leq L-1-\frac{2}{C'}$. The runtime for a 1D power-law $\mathcal{S}$-robust state transfer protocol with $\abs{\mathcal{S}}=2^k$ is lower bounded by 
\begin{align}
    t &\gtrsim 2 R(L) \frac{2^{(k-L+1)/(C'(L-k-1))}}{ \sqrt{C'(L-k-1)} C}\\
    &= R(L) \frac{2^{-1/C'}}{ \sqrt{L-k-1} \sqrt{C'} C}\\
    &\propto \frac{R(L)} { \sqrt{L-k-1}} \, .
\end{align}

Our bound can be parametrically stronger than previously known bounds and close to optimal in certain regimes.
Consider $k=L-h(L)$. Then, our bound is 
\begin{align}
    t \gtrsim \frac{R(L)} { \sqrt{h(L)}} \, .
\end{align}
For, say, $h(L)=\sqrt{L}$, in the regime $3/2<\alpha<5/2$, the new bound gives $t\gtrsim L^{\alpha-7/4}$ which is \emph{better} than the operator norm bound $t\gtrsim L^{\alpha-2}$ and the 2-norm bound $t\gtrsim 2^{-\sqrt{L}/2} L^{\alpha-1}$. Unfortunately, stemming from the fact that the p-norm light cone is not tight, especially as $p\rightarrow\infty$, this new bound is not always better than the operator-norm one. 
For $k=o(L)$, the new bound matches the operator-norm one for $2<\alpha<5/2$ and is worse for $5/2<\alpha<3$.
In this regime of $k$, $p_*\rightarrow\infty$ as $L\rightarrow \infty$.

\emph{Bridging protocol.}---We construct a fast, robust state transfer protocol for a specific subspace $\mathcal{S}$ and show that the runtime of this protocol almost matches Theorem~\ref{thmoptimalp}---highlighting the tightness of our Theorem and the optimality of the protocol. We explore additional protocols in the End Matter. Let $\ket{\psi}=a\ket{0}+b\ket{1}$. Suppose $L^\beta-1$ qubits at the left end of the chain and $L^\beta$ qubits at the right end are initialized to $\ket{0}$, while the $L-2L^\beta$ qubits in the middle are uninitialized as in Figure~\ref{fig:swapsetup}. The following protocol achieves state transfer on such ancilla states: 
\begin{figure*}[t]
    \centering
        \begin{tikzpicture}[scale=0.7]
\definecolor{softblue}{RGB}{110,155,205}
\definecolor{softred}{RGB}{215,115,115}
\definecolor{softgreen}{RGB}{120,190,145} 

\def\Rblue{0.48}

\tikzset{
    bluequbit/.style={
        fill=softblue,
        opacity=0.85
    }
}

\newcommand{\psiQubit}[2]{
    \path[fill=softgreen, opacity=0.85] (#1,#2) circle (\Rblue);

    \foreach \k in {1,...,12} {
        \pgfmathsetmacro{\rk}{\Rblue * \k / 12}
        \pgfmathsetmacro{\ok}{0.75 * exp(-0.25*\k)}
        \path[fill=white, opacity=\ok] (#1,#2) circle (\rk);
    }
    \node at (#1,#2) {$\ket{\psi}$};
}

\psiQubit{0}{0}

\foreach \i in {1,2} {
    \path[bluequbit] (\i*1.4,0) circle (\Rblue);
}

\node[opacity=0.5] at (4.4,0) {$\cdots$};

\foreach \i in {0,1,2,3,4} {
    \foreach \k in {1,...,12} {
        \pgfmathsetmacro{\rk}{\Rblue * \k / 12}
        \pgfmathsetmacro{\ok}{0.8 * exp(-0.15*\k)}
        \path[fill=softred, opacity=\ok]
            (6.2+\i*1.4,0) circle (\rk);
    }
}

\node[opacity=0.5] at (13.6,0) {$\cdots$};

\foreach \i in {0,1,2} {
    \path[bluequbit] (15.2+\i*1.4,0) circle (\Rblue);
}

\draw[
    decorate,
    decoration={brace, amplitude=6pt, mirror}
]
(-0.8,-1.0) -- (5,-1.0)
node[midway, below=6pt, opacity=0.8] {$L^\beta$};

\draw[
    decorate,
    decoration={brace, amplitude=6pt}
]
(-0.8,1.0) -- (18.8,1.0)
node[midway, above=6pt, opacity=0.8] {$L$};

\draw[
    decorate,
    decoration={brace, amplitude=6pt, mirror}
]
(13,-1.0) -- (18.8,-1.0)
node[midway, below=6pt, opacity=0.8] {$L^\beta$};

\end{tikzpicture}

\caption{The setup for the bridging protocol with qubits on each end of the chain initialized. Blue and red circles respectively indicate ancilla qubits in initialized and uninitialized states at $t=0$.}
    \label{fig:swapsetup}
\end{figure*}
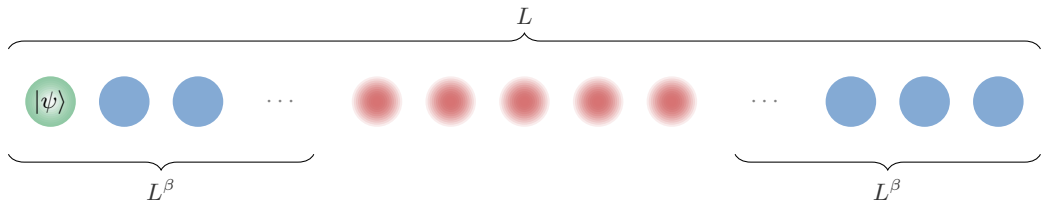

\begin{enumerate}
    \item On the left end of the chain, recursively build the GHZ state $a \ket{\bar{0}}+ b \ket{\bar{1}}$ following \cite{Tran2021a}. Simultaneously, on the right end build $ \frac{1}{\sqrt{2}} (\ket{\bar{0}}+ \ket{\bar{1}})$. Takes time $t_{GHZ}(L^\beta)$.
    \item Apply a direct controlled-phase between the GHZ states across the gap, 
    \begin{equation}
        H=L^{-\alpha}\sum_{k \in \rm left} \sum_{l \in \rm right} \dyad{1}_k \otimes \dyad{1}_l.
    \end{equation} At the end of this step, the left and right ends are in the state 
    $a \ket{\bar{0}} \otimes \frac{1}{\sqrt{2}} (\ket{\bar{0}}+ \ket{\bar{1}}) + b \ket{\bar{1}}\otimes  \frac{1}{\sqrt{2}} (\ket{\bar{0}}- \ket{\bar{1}})$. Takes time $t_\text{direct}$.
    \item Apply a generalized Hadamard to the right end of the chain, following \cite{Tran2021a}, to create the global GHZ state $a \ket{\bar{0}} \otimes\ket{\bar{0}}+  b \ket{\bar{1}} \otimes \ket{\bar{1}}$. Takes time $t'_{GHZ}(L^\beta)$. 
    \item Undo the previous steps on to the target site on the right end of the chain. Takes time $t_{GHZ}(L^\beta)+t'_{GHZ}(L^\beta)+t_\text{direct}$.
\end{enumerate}
The total runtime of this protocol is 
\begin{equation}
    \tau = 2(t_{GHZ}(L^\beta)+t'_{GHZ}(L^\beta) +t_\text{direct}).
\end{equation}

Consider $3/2<\alpha<2$. From \cite{Tran2021a}, $t'_\text{GHZ}(L^\beta)\sim t^{}_\text{GHZ}(L^\beta)\sim \ln^\kappa L^\beta \sim \ln^\kappa L$. Each direct coupling has strength bounded by $ L^{-\alpha}$, but there are $L^{2\beta}$ such couplings so $t_\text{direct}\sim L^{\alpha-2\beta}$. Thus for $\alpha>2\beta$, the total protocol time is dominated by the direct coupling step, $\tau \sim 2 L^{\alpha-2\beta}$.
In the same range of $\alpha$, from Theorem~\ref{thmoptimalp} we obtain the runtime lower bound of
\begin{equation}\label{gaplower}
    \tau \gtrsim L^{\alpha-3/2-\beta/2} \, ,
\end{equation}
which is exponentially better than the runtime bounds from both the operator and Frobenius light cones, which are $t\gtrsim \ln L$ and $t \gtrsim 2^{-L^\beta} L^{\alpha-1}$ respectively (see End Matter). In fact, our bound is within a small-degree polynomial factor of the protocol runtime. Supposing $\beta=1-\epsilon<\alpha/2$, the optimal time for this state transfer task is constrained by
\begin{equation}
L^{\alpha-2+\epsilon/2}  \lesssim \tau_\text{opt.} \lesssim L^{\alpha-2+2\epsilon},    
\end{equation}
where the lower bound comes from Theorem~\ref{thmoptimalp} and the upper bound from the runtime of the bridging protocol. For small $\epsilon$, the two almost coincide.

\emph{Outlook.}---In this work, we have initiated a study of the robustness of state transfer to changes in the ancilla state. Identifying the Schatten $p$-norms as a key technical tool, we prove a tight restriction on the growth of commutator $p$-norms under state transfer. In combination with existing light cones, these restrictions give us new and stronger lower bounds on the runtime of partially state-dependent state transfer protocols.

Our work raises several interesting questions. Consider the scenario where the ancillas are in thermal states. A state transfer protocol that works perfectly in this setting must be completely state-independent. However, we expect that at low temperatures a state-dependent protocol should work reasonably well. By considering \emph{approximate} state transfer, we expect to resolve this tension. Generalizing our results to this setting, and in particular developing the stabilizer picture for approximate transfer, is important future work.

Our results show that for any subspace of small dimension, there is a state transfer protocol that only grows small commutator norms. Thus, any light cone implies only a short minimum runtime for such a protocol. However, we expect that there are adversarial choices of subspaces that have small dimension but still slow down state transfer; for example, if the ancillas are highly entangled. 
Investigating this is an interesting future direction.

It would also be interesting to extend our analysis to the setting of state transfer with nonunitary dynamics, such as measurements and feedback, for which there exists a generalized Lieb-Robinson bound when one also counts the number of measurements as a resource \cite{friedman2023, Devulapalli_2024}. In this scenario, state transfer can be achieved in constant time with even local interactions using a quantum repeater. Nonetheless, its success requires both the careful creation of specially entangled resource states as well as precise knowledge of all measurement outcomes for the final feedback step. We provide some preliminary analyses in the SM for the scenario of measurements and feedback with imperfect local measurements. This extension could potentially be of interest in the resource analysis of fault-tolerant quantum repeaters \cite{Childress_2005, Raussendorf_2006, Wo_2023, Choe_2024}.

\emph{Acknowledgments.}---We thank Andrew Lucas, Nicole Yunger Halpern, and Peter Zoller for helpful discussions. 
T.U.\ acknowledges the support of the Natural Sciences and Engineering Research Council of Canada (NSERC) through the Doctoral Postgraduate Scholarship. We were  supported in part by the DoE ASCR Quantum Testbed Pathfinder program (award No.~DE-SC0024220 and No.~DE-SC0019040), NSF STAQ program, NSF QLCI (award No.~OMA-2120757), ONR MURI, AFOSR MURI,  DARPA SAVaNT ADVENT, DARPA Agreement HR00112490357, ARL (W911NF-24-2-0107) NQVL:QSTD:Pilot:DLPQC, and NQVL:QSTD:Pilot:FTL. We also acknowledge support from the U.S.~Department of Energy, Office of Science, Accelerated Research in Quantum Computing, Fundamental Algorithmic Research toward Quantum Utility (FAR-Qu)  and from the U.S.~Department of Energy, Office of Science, National Quantum Information Science Research Centers, Quantum Systems Accelerator (award No.~DE-SCL0000121).

\bibliography{mainbib}
\appendix 
\onecolumngrid
\center{\normalsize\textbf{End Matter}}
\vspace{1em}
\twocolumngrid
\justifying

\emph{Power-law Lieb-Robinson bounds and light cones}.---For ease of reference, we summarize some of the relevant light cones for power-law Hamiltonians. Let $r=\mathcal{D}(i,f)$.
Lieb-Robinson bounds are known in arbitrary dimension $d$.
From \cite{Kuwahara_2020, Tran2021lrb, Chen2023}, we have that
\begin{equation}\label{eq:power-law LR}
\norm{[U^\dagger X_f U,Z_i]  }_\infty \geq \delta \implies  t \gtrsim g(r,\delta), 
\end{equation}
where
\begin{equation}
    g(r,\delta)=
    \begin{cases}
    \alpha \ln r + \ln \delta  &\quad d<\alpha < 2d\\
    C r^\frac{\alpha-2d}{\beta} \delta^\frac{1}{\beta} &\quad 2d<\alpha <2d+1\\
    C (\delta(r- vt))^\frac{\alpha}{2d+1} &\quad \alpha>2d+1. 
    \end{cases}
\end{equation}
Here, $C,v,\beta$ are constants that depend only on $\alpha, d$.
By the monotonicity of $p$-norms, any Lieb-Robinson bound is also a light cone for the Frobenius norm. However, in 1D, an improved bound is known. From Theorem 4.1 of \cite{Chen2021} and Table I of \cite{Tran2021a}, we have that
\begin{equation}\label{eq:Frob bound}
\norm{[U^\dagger X_f U,Z_i]  }_2 \geq \delta \implies  t \gtrsim \delta^2 C g(r),  
\end{equation}
where 
\begin{equation}\label{eq:Frob g(r)}
    g(r)=
    \begin{cases}
    r^{\alpha-1} &\quad 1<\alpha < 2\\
    r/\ln^2(r) &\quad \alpha = 2\\
    r/ \ln(r) &\quad \alpha>2.    
    \end{cases}
\end{equation}
Both the operator and Frobenius bounds are tight asymptotically in $r$, i.e. there exist explicit Hamiltonians that saturate them up to subpolynomial corrections~\cite{Chen2021,Tran2021lrb}. In fact, in most ranges of $\alpha$, they are saturated by completely state-dependent and completely state-independent state transfer protocols, respectively.

\emph{Additional Robust State Transfer Protocols: Symmetric protocol}.---In this section, we extend the fast GHZ protocol in \cite{Tran2021a} to work for a larger class of ancilla states while retaining the same runtime scaling. In particular, we give an $\mathcal{S}$-robust state transfer protocol with $\mathcal{S}$ containing the symmetric subspace,
which is the subspace spanned by tensor product states of the form $\ket{\phi}^{\otimes L-1}$, known as independent and identically distributed states. 
(In fact we show the protocol we construct ends up working for an even larger subspace.) We emphasize that, by linearity, this protocol even achieves fast state transfer on certain mixed and entangled states of the ancilla qubits.

Our construction applies to qudits in arbitrary lattices, but we illustrate the point with qubits in 1D. Our strategy for the protocol is to first unitarily reset a subset of the sites to $\ket{0}$ and then use the fast GHZ protocol on this reset sublattice. The reset is accomplished by grouping neighbouring qubits in groups of $l$ and applying a unitary $U_\text{reset}$ on each group to reset the first qubit. That is, for all $\ket{\phi}$,
\begin{equation}
    U_\text{reset} \ket{\phi}^{\otimes l} = \ket{0}\otimes \ket{\phi'}.
\end{equation}
where $\ket{\phi} $ is supported on the remaining $l-1$ qubits. Since unitaries preserve inner products, such a $U_\text{reset}$ exists if and only if the Hilbert space of $l-1$ qubits is large enough. For $l$ qubits, the symmetric subspace has dimension $l+1$. The subspace dimension after resetting the first qubit is $2^{l-1}$. The smallest $l$ such that $2^{l-1}\geq l+1$ is $l=3$. For this $l$, a reset unitary can be explicitly defined as
\begin{equation}
U_{\text {reset}}: \begin{cases}
\ket{000} &\rightarrow \ket{000}\\
\ket{111} &\rightarrow \ket{011}\\
\frac{1}{\sqrt 3}(\ket{001} + \ket{010}+\ket{100}) &\rightarrow \ket{001}\\
\frac{1}{\sqrt 3}(\ket{011} + \ket{101}+\ket{110}) &\rightarrow \ket{010}\\
\frac{1}{\sqrt 6}(\ket{001} -2 \ket{010}+\ket{100}) &\rightarrow \ket{100}\\
\frac{1}{\sqrt 6}(\ket{011} -2 \ket{101}+\ket{110}) &\rightarrow \ket{101}\\
\frac{1}{\sqrt 2}(\ket{001} -\ket{100}) &\rightarrow \ket{110}\\
\frac{1}{\sqrt 2}(\ket{011} -\ket{110}) &\rightarrow \ket{111}
\end{cases}.
\end{equation}
Note that $U_{\text {reset}}$ sends all symmetric states (first four rows) to a state of the form $\ket{0} \otimes \ket{\phi'}$. For more general lattices one can round $l=3$ up to the size of a unit cell which tiles the lattice.

To summarize, our state transfer protocol is to first apply $U_{\text {reset}}$ to groups of 3 qubits and then do the original fast GHZ protocol on the sublattice that has been reset to $\ket{0}$. 
The sublattice's effective lattice spacing is changed by a constant. This doesn't impact the scaling of the fast GHZ protocol runtime. The initial reset unitary acts locally so can be performed in constant time. Hence, our protocol has the same runtime scaling as the original fast GHZ protocol but works for any ancilla state in the symmetric subspace.
In fact, it works for an even larger subspace than this. Since the reset unitary is applied independently to each group of 3 qubits, the protocol is $\mathcal{S}$-robust with $\mathcal{S}$ the subspace spanned by tensor product states where each set of 3 qubits is in a possibly different symmetric state. This subspace has dimension $4^{(L-1)/3}$, compared to the dimension $L$ of the symmetric subspace on $L-1$ qubits.

A similar state transfer protocol works when $\mathcal{S}$ is spanned by $m$ states, each a tensor product over single sites: $\{ \bigotimes_{j=1}^{L-1} \ket{\phi^i_j} \}_{i=1}^m$. We need to determine the smallest $l$ such that the reset can be achieved. Given $l$, consider each of the groups of $l$ qubits---without loss of generality we focus on the first group. Let $G$ be the Gram matrix of the ensemble $\left\{ \bigotimes_{j=1}^{l} \ket{\phi^i_j} \right\}_{i=1}^m$. A reset unitary on this group is guaranteed to exist as long as $2^{l-1}\geq \text{rank}(G)$. The smallest $l$ is the one such that this condition is satisfied for all $(L-1)/l$ groups. 
$\text{rank}(G)$ can be as large as $m$. As $m$ increases, so does the time for state transfer because performing the reset step may take longer and because the effective spacing $l$ of the reset sublattice grows.
One could consider even more general scenarios, e.g., by allowing $l$ to vary along the lattice as an adaptive parameter that depends on how badly initialized a patch of ancillas is. This could be of physical relevance in architectures where the locations of the noisiest qubits are characterized.

\begin{figure}
    \centering
    \includegraphics[width=\linewidth]{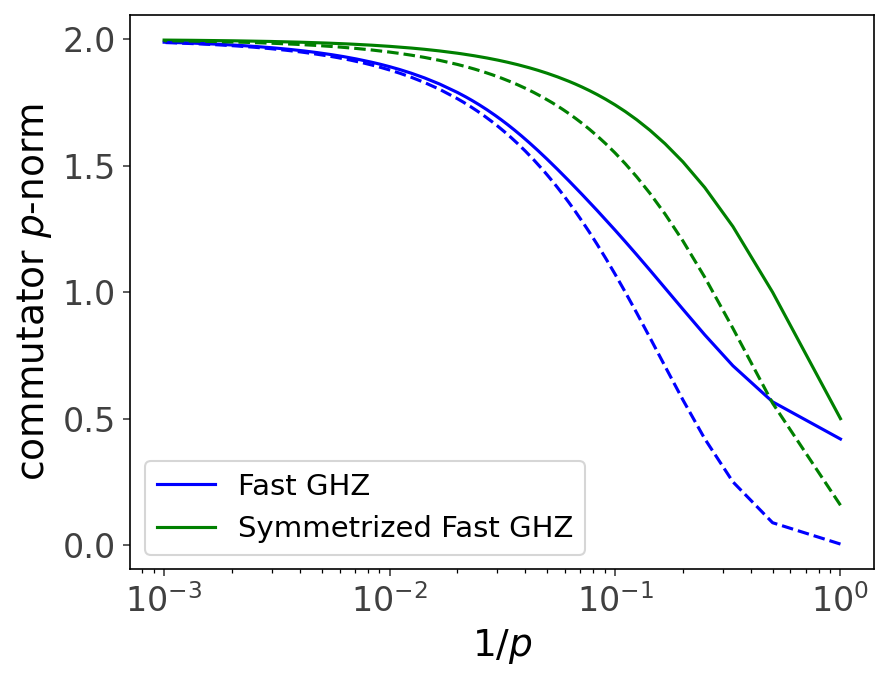}
    \caption{Commutator norms for fast GHZ protocol (blue) and its symmetrized version (green). Actual values (solid) compared to lower bounds (dashed). }
    \label{fig:thmbound}
\end{figure}

In Figure~\ref{fig:thmbound}, we plot numerically computed commutator norms for both the original and symmetrized protocols along with their lower bounds from Theorem~\ref{thmmainpnorm}. As expected, the symmetrized protocol grows a larger commutator and has a larger lower bound.
The detailed calculations are as follows: Numerically, we set $L=10$ (recall the number of ancilla qubits is $L-1$). For the unsymmetrized protocol, $k=0$. Hence, the lower bound is $2\cdot 2^{(1-L)/p}$. For the symmetrized protocol, the subspace dimension is $4^{(L-1)/3}$, i.e. $k= \frac{2}{3}(L-1)$, so the lower bound is $2\cdot 2^{-(L+1)/3p}$. For the unsymmetrized protocol we use direct controlled-NOTs between the first site and the four other sites on the left half, to build a GHZ state that encodes the unknown state $ a\ket{\bar{0}}+b\ket{\bar{1}}$. Simultaneously, on the right hand side we apply a Hadamard on the last site and then do the same coupling to build a 5-qubit GHZ state $\frac{1}{\sqrt{2}} (\ket{\bar{0}}+\ket{\bar{1}})$. Following~\cite{Tran2021a} and as summarized in the main text, we apply controlled phases between all sites on the left and right sides, do generalized Hadamards, and disentangle on to the right side of the chain, moving the unknown state onto the tenth qubit.

For the symmetrized version, we first apply $U_\text{reset}^{\otimes 3}$, with the convention that it resets the rightmost of the three qubits, to leave the fourth, seventh, and tenth ancilla qubits in $\ket{0}$. After a Hadamard on the last qubit, and direct controlled-NOTs to build the initial GHZ state, we proceed with the usual fast GHZ protocol on this four-qubit sublattice, achieving state transfer from the first to tenth qubit.

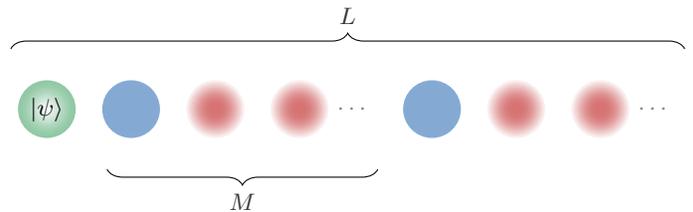
\begin{figure}[t]
    \centering
        \begin{tikzpicture}[scale=0.8]

\definecolor{softblue}{RGB}{110,155,205}
\definecolor{softred}{RGB}{215,115,115}
\definecolor{softgreen}{RGB}{120,190,145} 
\def\Rblue{0.48}

\tikzset{
    bluequbit/.style={
        fill=softblue,
        opacity=0.85
    }
}

\newcommand{\psiQubit}[2]{
    \path[fill=softgreen, opacity=0.85] (#1,#2) circle (\Rblue);

    \foreach \k in {1,...,12} {
        \pgfmathsetmacro{\rk}{\Rblue * \k / 12}
        \pgfmathsetmacro{\ok}{0.75 * exp(-0.25*\k)}
        \path[fill=white, opacity=\ok] (#1,#2) circle (\rk);
    }

    \node at (#1,#2) {$\ket{\psi}$};
}
\psiQubit{-1.4}{0}

\path[bluequbit] (0,0) circle (\Rblue);

\foreach \i in {1,2} {
    \foreach \k in {1,...,12} {
        \pgfmathsetmacro{\rk}{\Rblue * \k / 12}
         \pgfmathsetmacro{\ok}{0.8 * exp(-0.15*\k)}
        \path[fill=softred, opacity=\ok]
            (\i*1.4,0) circle (\rk);
    }
}

\node[opacity=0.5] at (3.7,0) {$\cdots$};

\path[bluequbit] (5,0) circle (\Rblue);

\foreach \i in {1,2} {
    \foreach \k in {1,...,12} {
        \pgfmathsetmacro{\rk}{\Rblue * \k / 12}
         \pgfmathsetmacro{\ok}{0.8 * exp(-0.15*\k)}
        \path[fill=softred, opacity=\ok]
            (5+\i*1.4,0) circle (\rk);
    }
}

\node[opacity=0.5] at (8.7,0) {$\cdots$};

\draw[
    decorate,
    decoration={brace, amplitude=6pt, mirror}
]
(-0.4,-1.0) -- (4.1,-1.0)
node[midway, below=6pt, opacity=0.8] {$M$};

\draw[
    decorate,
    decoration={brace, amplitude=6pt}
]
(-2,1.0) -- (9.2,1.0)
node[midway, above=6pt, opacity=0.8] {$L$};

\end{tikzpicture}
\caption{The setup with 1 in every $M$ qubits initialized. Blue and red circles respectively indicate ancilla qubits in initialized and uninitialized states at $t=0$.}
    \label{fig:subsetup}
\end{figure}

\emph{Additional Robust State Transfer Protocols: Sublattice protocol}.---In this section, we generalize from the setup of the symmetric protocol to a scenario where 1 in every $M$ qubits is in $\ket{0}$ and the remaining qubits are in $\frac{1}{2}\id$ as in Figure~\ref{fig:subsetup}. Then we use the fast GHZ protocol of \cite{Tran2021a} on the initialized sublattice. Working on the sublattice is equivalent to increasing the original lattice spacing by a factor of $M$. For power-law Hamiltonians, this means each interaction term decays by a factor of $\frac{1}{M^\alpha}$, so each unitary takes time at most $M^\alpha$ as long. Hence, the overall protocol runtime just gains a constant prefactor of $M^\alpha$. For example, in the regime $3/2<\alpha<2$, this protocol takes time 
\begin{equation}
    t\approx M^\alpha (\log L)^{\kappa_\alpha}.
\end{equation}
In comparison, the lower bound from Thm.~\ref{thmoptimalp} is
\begin{equation}\label{lowerboundM}
    t\gtrsim \sqrt{M} L^{\alpha-2}.
\end{equation}
This suggests that either a tighter bound or a faster state transfer protocol exists.

We can compare to the setup of the bridging protocol in the main text where all the uninitialized qubits were in the middle of the chain. For a fair comparison, assume $L/2M$ qubits at each end of the chain are initialized. Since Thm.~\ref{thmmainpnorm} only depends on the size of the subspace $\mathcal{S}$, the lower bounds for both setups, Eqs.~\eqref{gaplower} and \eqref{lowerboundM}, are the same. However, the bridging protocol has a faster runtime than the sublattice one. 
The locations of the $L/M$ initialized qubits are very different in the two cases.

Theorem~\ref{thmmainpnorm} is tight---for any $\mathcal{S}$ there is a state transfer protocol that saturates it. But the bound only depends on the dimension of $\mathcal{S}$. Hence, two protocols for two different subspaces with the same dimension will have the same lower bound on their commutator norm and thus on their runtime. One possibility, which seems unlikely, is that optimal times for $\mathcal{S}$-robust state transfer only depend on $\abs{\mathcal{S}}$. The other possibility is that commutator-norm-based bounds are insufficient for tightly constraining partially-state-dependent state transfer. 
Thus, to obtain stronger bounds on state transfer times, one has to go beyond the eigenvalues of the commutators and take into account the structure of the eigenvectors as well.

\end{document}

\onecolumngrid
\newpage

\setcounter{secnumdepth}{1}
\setcounter{section}{0}
\renewcommand{\thesection}{S\arabic{section}}
\setcounter{thm}{0}
\renewcommand{\thethm}{S\arabic{thm}}
\renewcommand{\thecor}{S\arabic{cor}}
\setcounter{lemma}{0}
\renewcommand{\thelemma}{S\arabic{lemma}}
\setcounter{equation}{0}
\renewcommand{\theequation}{S\arabic{equation}}
\setcounter{table}{0}
\renewcommand{\thetable}{S\arabic{table}}
\setcounter{figure}{0}
\renewcommand{\thefigure}{S\arabic{figure}}
\setcounter{defi}{0}
\renewcommand{\thedefi}{S\arabic{defi}}

\renewcommand{\bibnumfmt}[1]{[S#1]}
\renewcommand{\citenumfont}[1]{S#1}

\newtheorem*{thm1}{Theorem 1}

\title{Supplemental Material for ``Tight Robustness-Runtime Tradeoff for Quantum State Transfer''}
\maketitle
 \raggedright
\setlength{\parindent}{20pt}
 \justifying

\newcommand{\yifan}[1]{{\color{cyan} Yifan:~#1}}
\newcommand{\tw}[1]{{\color{Rhodamine} Twesh:~#1}}
\newcommand{\connor}[1]{{\color{orange} Connor:~#1}}

In this Supplemental Material, we provide more details about our setup, proofs of the two theorems in the main text, as well as some extensions of the results presented in the main text. In \cref{sec:defi}, we elaborate on and motivate our definition of state transfer, especially our notion of robustness. In \cref{sec:noisy state transfer}, we build up the machinery to prove the main text's Thm.~1 as well as a more detailed treatment of the protocol which saturates it. In \cref{sec:proofthm2}, we prove the main text's Thm.~2, and \cref{sec:qudit} gives the details of extending the main text's results to qudits with higher local Hilbert space dimension. Finally, in \cref{sec:measurements}, we extend our results to non-unitary state transfer protocols, notably those including mid-circuit measurements.

\section{Definition of State Transfer}\label{sec:defi}
Here we motivate and present in detail our definition of robust state transfer. We start with the following definition of state transfer.
A state transfer protocol is a Hamiltonian that generates a unitary $U$ which transports an unknown state on site $i$ to $f$. That is, a state transfer protocol working for the ancilla state $\sigma_\text{anc}$ satisfies
\begin{equation}
    U (\dyad{\psi}_i \otimes \sigma_\text{anc})U^\dagger  = \sigma'_{\rm anc'} \otimes \dyad{\psi}_f \quad \forall \ket{\psi},
\end{equation}
where $\sigma'_{\rm anc'}$ can be arbitrary. As a consequence of this definition, $\sigma'$ must be independent of the transferred state $\ket{\psi}$.
Often, the unitary is implemented via a time-dependent, local Hamiltonian, where the precise notion of locality depends on the setup.

Our goal is to study the role of the ancilla state in state transfer. We hence need to define precisely the set of ancilla states for which a given protocol works. To this end, we first prove the important fact that a state transfer protocol works for every one of a set of ancilla states $\{\ket{\Phi_k}\}$ if and only if it works for any one strict convex combination $\sum_k \lambda_k \dyad{\Phi_k}$.
The forward direction is immediate by linearity. To prove the converse, let $U (\ket{\psi}\otimes \ket{\Phi_k}) = \ket{\chi_k}$. By assumption,
\begin{align}
\sum_k \lambda_k \Tr_{\rm anc'} (\dyad{\chi_k}) = \dyad{\psi}.
\end{align}
Each term $\lambda_k \Tr_{\rm anc'} (\dyad{\chi_k})$ is a nonzero positive operator. Since these operators sum to a rank-1 operator, each must be proportional to that rank-1 operator. Therefore,
\begin{equation}
 \Tr_{\rm anc'} (\dyad{\chi_k})=\dyad{\psi} \quad \forall k .
\end{equation}
Each $\ket{\chi_k}$ admits a Schmidt decomposition between site $f$ and the rest of the sites: $\ket{\chi_k}=\sum_j c_j \ket{a_j}_{\rm anc'} \otimes  \ket{b_j}_f$. So, we have 
\begin{align}
 \Tr_{\rm anc'} (\dyad{\chi_k}) = \sum_j \abs{c_j}^2 \dyad{b_j}=\dyad{\psi}.
\end{align}
By the same reasoning as above, each $\ket{b_j}=\ket{\psi}$. Therefore, $\ket{\chi_k}=\ket{\chi'_k}\otimes \ket{\psi}$: the state transfer protocol works for each state in the convex combination. As a special case, we recover the result known in lore that a state transfer protocol works for every ancilla state if and only if it works for the maximally mixed ancilla state. 

By linearity, the set of ancilla states for which a state transfer protocol works is a convex set. In light of the above fact, the extreme points of this set are pure states. Thus, it is natural to consider the \emph{subspace} spanned by these pure states. This motivates our definition of $\mathcal{S}$-robust state transfer.

We could also consider the transfer of more than one qubit. The above definitions generalize by simply replacing $\dyad{\psi}_i$ with a multipartite state $\dyad{\psi}_{i^1,...,i^n}$. In the main text, we discuss an equivalent description of state transfer in the Heisenberg picture. This generalizes to the multiqubit transfer scenario. $U$ achieves multiqubit $\mathcal{S}$-robust state transfer if and only if
\begin{align}\label{stabilizermulti}
U^\dagger X_{f_k} U &=X_{i_k} S_x^k \quad \text{and} \\
U^\dagger Z_{f_k} U &=Z_{i_k} S_z^k,
\end{align}
where $S^k_{x,z}$ are unitaries that obey 
\begin{equation}\label{subspacemulti}
    S_x^k (\ket{\psi}_{i^1 ... i^n} \otimes \ket{\Phi}_{\rm anc})  = S_z^k (\ket{\psi}_{i^1 ... i^n} \otimes \ket{\Phi}_{\rm anc}) = \ket{\psi}_{i^1 ... i^n} \otimes \ket{\Phi}_{\rm anc} 
\end{equation} for all $\ket{\Phi}\in\mathcal{S}$ and $\ket{\psi}\in (\mathbb{C}^{D})^{\otimes{n}}$.

\section{State transfer with noisy ancillas} \label{sec:noisy state transfer}
In this section, we provide the proof of Thm.~1 connecting the robustness of state transfer to the growth of commutator norms. We also give an explicit protocol demonstrating that our theorem is tight.

\subsection{Commutator norm bounds}
We begin by restating Eq.~(12) from the main text:\begin{equation}\label{maincommutator}
    [U^\dagger X_f U,Z_i] =[X_i S_x,Z_i].
\end{equation}
In order to prove our main result, we first obtain some intermediate facts. 

Define $V\equiv Y_i S_x$.
Our first lemma relates the commutator in Eq.~\eqref{maincommutator} to the Hermitian part of $V$.
\begin{lem}\label{realpart}
    Let $V= Y_i S_x$ with eigendecomposition $V= \sum_k e^{i \theta_k} \dyad{\chi_k}$. Then,
    \begin{equation}
        [X_i S_x, Z_i]=-2i \He V = -2i \sum_k \cos \theta_k \dyad{\chi_k}.
    \end{equation}
\end{lem}
\begin{proof}
We have 
\begin{align}
[X_i S_x, Z_i] &=S^\dagger_x X_i Z_i - Z_i X_i S_x\\
&=-i S^\dagger_x Y_i - i Y_i S_x\\
&= -i( V^\dagger+V)\\    
&= -2i \He V\\    
&= -2i \sum_k \cos \theta_k \dyad{\chi_k},  
\end{align}
where in the first line we have used Eq.~(6) of the main text.
\end{proof}
Two useful corollaries now follow. The first concerns shared eigenvectors of the commutator and $V$ while the second relates their $p$-norms.
\begin{cor}\label{corshared}
A state is a $+1$ ($-1$) eigenstate of $V$ iff it is a $-2i$ ($+2i$) eigenvector of $[X_i S_x, Z_i]$.  
\end{cor}
\begin{proof}
This follows from Lemma \ref{realpart} and the fact that $\He V$ and $V$ agree on their real eigenspaces.
\end{proof}

\begin{cor}\label{corpnorm}
The commutator norm is
$\norm{[X_i S_x, Z_i]}_p =2 \norm{ \He V}_p= 2 \left(\frac{\sum_k \abs{\cos \theta_k}^p }{2^L}\right)^\frac{1}{p} $.
\end{cor}
\begin{proof}
The result follows from taking the $p$-norm on both sides of the equation in Lemma \ref{realpart} and using the positive homogeneity of this norm.
\end{proof}

We next prove some results linking $V$ to the properties of the underlying state transfer protocol. Let $\ket{R}$ and $\ket{L}$ denote the $+1$ and $-1$ eigenstates of $Y$, respectively.

\begin{lem}\label{thmSandV}
    Let $\ket{\Phi}$ be an ancilla state for which state transfer works. $\ket{R}\otimes\ket{\Phi} $ is a $+1$ eigenvector of $S_x$ if and only if it is a $+1$ eigenvector of $V$. Similarly, $\ket{L}\otimes\ket{\Phi} $ is a $+1$ eigenvector of $S_x$ if and only if it is a $-1$ eigenvector of $V$.
\end{lem}
\begin{proof}
    Assuming $\ket{R/L}\otimes\ket{\Phi}$ are $+1$ eigenvectors of $S_x$, it is easily verified that 
    \begin{align}
        V \left(\ket{R/L}\otimes\ket{\Phi}\right) &= Y_i \ket{R/L}\otimes\ket{\Phi}\\
        &=\pm \ket{R/L}\otimes\ket{\Phi}.
    \end{align}

For the converse direction, we use the unitarity of $V$ and $S_x$, 
    \begin{align}
    V \ket{R/L}\otimes\ket{\Phi} = \pm \ket{R/L}\otimes\ket{\Phi} &\implies  V^\dagger \ket{R/L}\otimes\ket{\Phi} = \pm \ket{R/L}\otimes\ket{\Phi} \\
    & \implies S_x^\dagger Y_i \ket{R/L}\otimes\ket{\Phi} = \pm \ket{R/L}\otimes\ket{\Phi}\\
        & \implies S_x^\dagger \left( \ket{R/L}\otimes\ket{\Phi}\right) = \ket{R/L}\otimes\ket{\Phi}\\
        & \implies S_x \left( \ket{R/L}\otimes\ket{\Phi} \right) = \ket{R/L}\otimes\ket{\Phi}.
    \end{align}
\end{proof}
Note that this lemma is not simply the statement that $[Y_i, S_x]=0$ as in general the two do not commute.
Recalling our earlier notation, let $\mathcal{S}_x$ denote the subspace of ancilla states $\ket{\Phi}$ such that $\ket{\psi}\otimes\ket{\Phi}$ is a $+1$ eigenvector of $S_x$ (but not necessarily of $S_z$).
\begin{cor}\label{cordimension}
     The $p$-norm of $\Re V$ is related to the subspace dimension via $\norm{\Re V}_p \geq 2^{(1-L)/p}\cdot \abs{\mathcal{S}_x}^{1/p}$.
\end{cor}
\begin{proof}
    Let $\{\ket{\Phi_j}\}$ be a basis for $\mathcal{S}_x$. By definition, $\ket{R}\otimes\ket{\Phi_j}$ and $\ket{L}\otimes\ket{\Phi_j}$ are $+1$ eigenstates of $S_x$. By Lemma \ref{thmSandV}, they are then $\pm1$ eigenstates of $V$. Since Lemma \ref{thmSandV} applies to every $\ket{\Phi_j}$, $V$ and in turn $\He V$ have at least $\abs{\mathcal{S}_x}$ eigenvalues that are 1 and the same number that are $-1$. Thus, 
    \begin{align}
     \norm{\He V}_p&= 2^{-L/p} \Big(\sum_{\{\lambda\}} \abs{\Re\lambda}^p \Big)^{1/p}\\
     &\geq2^{-L/p} (2 \abs{\mathcal{S}_x})^{1/p}\\
     &=2^{(1-L)/p} \abs{\mathcal{S}_x}^{1/p}   .
    \end{align}
\end{proof}
Note the maximum value of $\abs{\mathcal{S}_x}$ is $\dim\mathcal{H}_{L-1} = 2^{L-1}$ which saturates the norm. The norm is also saturated when $V$ only has real eigenvalues, which will be relevant in the next section where we derive a saturating protocol. We are now in a position to state our main theorem connecting the robustness of state transfer protocols to commutator $p$-norms.
\begin{thm1}\label{thm:mainpnorm}
    Any $\mathcal{S}$-robust state transfer protocol satisfies $\norm{[X_i S_x, Z_i]}_p \geq 2\cdot 2^{(-L+1)/p} \abs{\mathcal{S}}^{1/p} $.
\end{thm1}
\begin{proof}
    First note that $\abs{\mathcal{S}}\leq \abs{\mathcal{S}_x}$. We have that
    \begin{align}
    \norm{[X_i S_x, Z_i]}_p &=2 \norm{ \He V}_p\\
    &\geq 2\cdot 2^{(1-L)/p}\cdot \abs{\mathcal{S}_x}^{1/p} \\
    &\geq 2\cdot 2^{(-L+1)/p} \abs{\mathcal{S}}^{1/p},
    \end{align}
    where the first line follows by Cor. \ref{corpnorm}, the second by Cor. \ref{cordimension}, and the third by assumption.
\end{proof}

\subsection{Optimal protocol}
We now verify that the state transfer protocol presented in the main text saturates Thm~1. Recall the unitary as defined in Eq.~(13) of the main text is
\begin{equation}
    U= \mathrm{SWAP}_{i,f} \otimes \Pi_\mathcal{S'}  +  (\mathrm{SWAP}_{i,f} \tilde{H}_i) \otimes \bar{\Pi}_\mathcal{S'}\, ,
\end{equation}
where the subspace projector is of the form $\Pi_\mathcal{S}=\Pi_\mathcal{S'} \otimes \id_f$, so that $\abs{\mathcal{S}}=2\abs{\mathcal{S'}}$.
Explicit computation gives
\begin{align}
    U^\dagger X_f U &= \left( \mathrm{SWAP}_{i,f} \otimes \Pi_\mathcal{S'}  +  \tilde{H}_i \mathrm{SWAP}_{i,f} \otimes \bar{\Pi}_\mathcal{S}\right) X_f \left( \mathrm{SWAP}_{i,f} \otimes \Pi_\mathcal{S'}  +  \mathrm{SWAP}_{i,f} \tilde{H}_i \otimes \bar{\Pi}_\mathcal{S'}\right)\\
    &=X_i \otimes \Pi_\mathcal{S'} \otimes \id_f +  Z_i \otimes \bar{\Pi}_\mathcal{S'} \otimes \id_f \, .
\end{align}
From this and a similar computation for $Z_f$, we can read off the stabilizer unitaries
\begin{align}\label{sxform}
S_x&=\id_i \otimes \Pi_\mathcal{S'} \otimes \id_f -i Y_i \otimes \bar{\Pi}'_\mathcal{S'} \otimes \id_f,  \\
S_z&=\id_i \otimes \Pi_\mathcal{S'} \otimes \id_f +i Y_i \otimes \bar{\Pi}'_\mathcal{S'} \otimes \id_f \, .
\end{align}
Note the sign difference between the first and second terms for $S_x$ versus $S_z$. From the projectors, it is clear that $S_x$ and $S_z$ share the same stabilizer subspace of dimension $2\abs{\mathcal{S}'}$. From Thm.~1, the lower bound on the commutator is thus $\norm{[X_i S_x, Z_i]}_p \geq 2\cdot 2^{(-L+2)/p} \abs{\mathcal{S'}}^{1/p}$. On the other hand, we can explicitly compute the commutator norm,
\begin{align}
    \norm{[X_i S_x, Z_i]}_p &=2 \norm{ \Re (Y_i S_x)}_p\\
    &= 2 \norm{\Re \big( Y_i \otimes \Pi_\mathcal{S'} \otimes \id_f -i \id_i \otimes \bar{\Pi}'_\mathcal{S'} \otimes \id_f   \big)  }_p\\
    &=2 \norm{ Y_i \otimes \Pi_\mathcal{S'} \otimes \id_f }_p\\
    &= 2 \cdot 2^{(2-L)/p}\abs{\mathcal{S'}}^{1/p},
\end{align}
where the first line follows from Corollary \ref{corpnorm} and the second line from substituting in Eq. \eqref{sxform}. This exactly matches the lower bound from Thm.~1 with $\Pi_\mathcal{S}=\Pi_\mathcal{S'} \otimes \id_f$. 

\section{Proof of Theorem 2}\label{sec:proofthm2}
Before proceeding to the proof, we motivate why it is important to consider different norms for different state transfer protocols.
Prior to this work, for a \emph{partially} state-dependent state transfer protocol, the only known bounds came from the operator norm, as Frobenius norms constrained only the fully state-independent case. Even for an almost state-independent protocol, the Frobenius-norm bound could not be applied. This is resolved by our Thm.~1, which allows any norm bound to be applied to any $\mathcal{S}$-robust state transfer protocol by taking the dimension of $\mathcal{S}$ into account. 
However, some such bounds are quite poor. First, consider a $\mathcal{S}$-robust state transfer protocol with $\abs{\mathcal{S}}=1$. Thm.~1 and Lieb-Robinson bounds [Eq.~(23) in the End Matter] imply the following minimum times for a power-law implementation of this protocol:
\begin{equation}
    t\gtrsim 
    \begin{cases}
    \ln L \quad 1<\alpha < 2\\
    L^{\alpha-2}  \quad      2<\alpha <3\\
    L \quad \alpha>3  .  
    \end{cases}
\end{equation}
On the other hand, Thm.~1 and the Frobenius light cones [Eq.~(25) in the End Matter] give
\begin{equation}
    t\gtrsim 2^{(1-L)/2} C
    \begin{cases}
    L^{\alpha-1} \quad 1<\alpha < 2\\
    L/\ln^2(L) \quad \alpha = 2\\
    L \quad \alpha>2  .  
    \end{cases}
\end{equation}
which are trivial bounds asymptotically in $L$. Hence for this particular $\abs{\mathcal{S}}$, operator norm light cones provide stronger runtime constraints.
In the other limit, consider state-independent state transfer, which corresponds to $\abs{\mathcal{S}}=2^{L-1}$. The operator-norm bounds still apply and give
\begin{equation}
    t\gtrsim 
    \begin{cases}
    \ln L \quad 1<\alpha < 2\\
    L^{\alpha-2}  \quad      2<\alpha <3\\
    L \quad \alpha>3    ,
    \end{cases}
\end{equation}
while the Frobenius bounds are
\begin{equation}
    t\gtrsim C
    \begin{cases}
    L^{\alpha-1} \quad 1<\alpha < 2\\
    L/\ln^2(L) \quad \alpha = 2\\
    L \quad \alpha>2    ,
    \end{cases}
\end{equation}
which are better for all $\alpha$. 
For $1<\alpha<2$, there is a crossover at $2^{1-L}\abs{\mathcal{S}}= L^{2-2\alpha} \ln L$ where both bounds have the same scaling.

We observe that $p=\infty$ bounds are tight for completely state-dependent state transfer, while $p=2$ are tight for completely state-independent (see also Figure~2 in the main text). On the other hand, both bounds are poor when applied to the other scenario. These observations are generalized by our Thm.~2: for a $\mathcal{S}$-robust state transfer protocol in 1D with a power-law Hamiltonian, the best $p$-norm light cone is from a value of $p$ that decreases with $\abs{\mathcal{S}}$.  

\begin{proof}
    Maximizing $2 R(L) \frac{2^{(k-L+1)/p}}{ \sqrt{p} C}$ over $p\geq 2$ is equivalent to maximizing $\frac{2^{(k-L+1)/p}}{ \sqrt{p}}=2^{{(k-L+1)/p}-\frac{1}{2 \ln 2} \ln p}$ and in turn to maximizing $h(p)={\frac{(k-L+1)}{p}-\frac{1}{2 \ln 2} \ln p}$. If $k-L+1=0$, then the maximum is clearly at $p=2$. For $k-L+1< 0$, the function tends to $-\infty$ as $p$ tends to $\infty$, so the maximum is located either at $p=2$ or where the derivative is zero. The derivative is 
\begin{align}
    \frac{d}{d p} \left[ {\frac{(k-L+1)}{p}-\frac{1}{2\ln 2} \ln p}\right] = \frac{L-k-1}{p^2} - \frac{1}{(2  \ln2) p } ,
\end{align}
and setting it to zero gives
\begin{equation}
    \frac{L-k-1}{p_c^2} - \frac{1}{(2  \ln 2) p_c  } =0 \iff \\
    \frac{L-k-1}{p_c} =\frac{1}{2\ln 2} \iff\\
    p_c = (2\ln 2) (L-k-1). 
\end{equation}
We have $h(2)=\frac{k-L+1}{2}-\frac{1}{2}=-\frac{1}{4\ln 2} p_c-\frac{1}{2}$ and $h(p_c)= -\frac{1}{2\ln 2}(1+ \ln p_c)$; the latter is larger.

\end{proof}

\section{Qudit state transfer}\label{sec:qudit}
In this section, we briefly discuss the extension of our results to qudits. Let $D$ be the qudit dimension. In this section, $X$ and $Z$ refer to the generalized Pauli matrices, also known as shift and clock operators, and $\omega=e^\frac{2 \pi i }{D}$. Recall that the commutator of an operator on the initial site $A_i$ with the time evolution of an operator on the final site $U^\dagger B_f U$ gives a necessary condition for state transfer. However, if $A_i$ and $U^\dagger B_f U$ commute, this constraint is trivial. For qudits, the commutator of $X$ and $Z$ is $[X,Z]=(1-\omega) X Z$; which tends to zero for large $D$. Thus, although the choice of $X$ and $Z$ works for qubits, it is poor for qudits. We propose a better choice for even and odd $D$ and then generalize our results for qubits. 

For even $D$, we consider the pair of operators $\xt=X^{D/2}$ and $Z$.
Recall that for qudits, necessary and sufficient conditions for state transfer are $U^\dagger X_f U=X_i S_x$ and $U^\dagger Z_f U=Z_i S_z$ where $S_x, S_z$ are stabilizer unitaries (i.e. states for which state transfer works are +1 eigenvectors of these unitaries). As $X$ and $Z$ generate a complete basis, it follows that any unitary, in particular $\xt$, transforms similarly: $U^\dagger \xt_f U = \xt_i \stx$ for some stabilizer unitary $\stx$. Recalling our earlier notation, let $\mathcal{\tilde{S}}_x$ denote the subspace of ancilla states $\ket{\Phi}$ such that $\ket{\psi}\otimes\ket{\Phi}$ is a $+1$ eigenvector of $\stx$.
Define the unitary $\tilde{Y}=\xt Z$ with eigendecomposition $\tilde{Y} = \sum \mu_k \dyad{e_k}$. 
As before, we have
\begin{align}
    [U^\dagger \xt_f U,Z_i] &= [\xt_i S_x,Z_i] \\
    &=  \stx^\dagger \xt_i  Z_i - Z_i \xt_i \stx\\
    &=  \stx^\dagger \xt_i  Z_i + \xt_i Z_i \stx\\
    &=  \stx^\dagger \yt_i + \yt_i \stx .
\end{align}

Let $\{\ket{\Phi_j}\}$ be a basis for $\mathcal{S}_x$. By definition, for all $k,j$, $\ket{e_k}\otimes\ket{\Phi_j}$ is a $+1$ eigenstate of $\stx$, and, as before, of $\stx^\dagger$. Then,
    \begin{align}
     \stx^\dagger \tilde{Y}_i \ket{e_k}\otimes\ket{\Phi_j}  &= \mu_k \stx^\dagger \ket{e_k}\otimes\ket{\Phi_j}\\
    &= \mu_k \ket{e_k}\otimes\ket{\Phi_j}.
    \end{align}
    Similarly, 
    \begin{align}
     \bra{e_k}\otimes\bra{\Phi_j} \tilde{Y}_i \stx  &= \mu_k \bra{e_k}\otimes\bra{\Phi_j} \stx \\
    &= \mu_k \bra{e_k}\otimes\bra{\Phi_j}. \label{mueval}
    \end{align}
    Taking the inner product with $\ket{e_k}\otimes\ket{\Phi_j}$ on both sides and taking absolute values gives 
    \begin{equation}
        \abs{\bra{e_k}\otimes\bra{\Phi_j} \tilde{Y}_i \stx \ket{e_k}\otimes\ket{\Phi_j}} = 1.
    \end{equation}
    By the Cauchy-Schwarz inequality, $\ket{e_k}\otimes\ket{\Phi_j}$ must be an eigenvector of $\tilde{Y}_i \stx$. By Eq.~\eqref{mueval} its eigenvalue must be $\mu_k$, 
    which also implies that $\ket{e_k}\otimes\ket{\Phi_j}$ is a right $\mu_k$ eigenvector of $\tilde{Y}_i \stx$.
    Thus, $\ket{e_k}\otimes\ket{\Phi_j}$ is a $2\mu_k$ eigenvector of $\stx^\dagger \yt_i + \yt_i \stx$.
    This means that for each $k$, the commutator has at least $\abs{\mathcal{S}_x}$ eigenvalues that are $2 \mu_k$. Thus, 
    \begin{align}
     \norm{\stx^\dagger \yt_i + \yt_i \stx}_p&= D^{-L/p} \Big(\sum_{\{\lambda\}} \abs{\lambda}^p \Big)^{1/p}\\
     &\geq D^{-L/p} \Big( \abs{\mathcal{S}_x} \sum_{k} \abs{2\mu_k}^p \Big)^{1/p} \label{eq_mu_sum} \\
     &\geq 2\cdot D^{(-L+1)/p} \abs{\mathcal{S}_x}^{1/p}.
    \end{align}
Then, following similar reasoning to the qubit case, we have that any qudit $\mathcal{S}$-robust state transfer protocol with even $D$ satisfies 
\begin{equation}
\norm{[\xt_i S_x, Z_i]}_p \geq 2\cdot D^{(-L+1)/p} \abs{\mathcal{S}}^{1/p}.    
\end{equation}

For odd $D$, it is not possible for two full-rank, single-site operators to anticommute. This prohibits an analogous construction to the $\tilde{Y}$ above. However, one can still obtain necessary conditions on state transfer by choosing the single-site operators to be the qudit Paulis $\tilde{X}$ and $Z$ of dimension $D-1$, embedded into a subspace of the larger Hilbert space. All steps above go through as before, except that in Eq.~\eqref{eq_mu_sum} one of the $\mu_k$ is now zero. Hence, the final bound for odd $D$ reads
\begin{equation}
\norm{[\xt_i S_x, Z_i]}_p \geq 2\cdot (D-1)^{1/p} D^{-L/p} \abs{\mathcal{S}}^{1/p}.    
\end{equation}

A protocol construction similar to the qubit case shows our qudit result is also tight. Let $\tilde{H}$ denote the qudit Hadamard or Fourier transform, which is the unitary that converts between the bases of $X$ and $Z$: $\tilde{H} Z \tilde{H}^\dagger=X$. Consider
\begin{equation}
    U= \mathrm{SWAP}_{i,f} \otimes \Pi_\mathcal{S'}  +  (\mathrm{SWAP}_{i,f} \tilde{H}_i) \otimes \bar{\Pi}_\mathcal{S'}\, . 
\end{equation}
Explicit computation gives
\begin{align}
    U^\dagger \xt_f U &= \left( \mathrm{SWAP}_{i,f} \otimes \Pi_\mathcal{S'}  +  \tilde{H}^\dagger_i \mathrm{SWAP}_{i,f} \otimes \bar{\Pi}_\mathcal{S}\right) \xt_f \left( \mathrm{SWAP}_{i,f} \otimes \Pi_\mathcal{S'}  +  \mathrm{SWAP}_{i,f} \tilde{H}_i \otimes \bar{\Pi}_\mathcal{S'}\right)\\
    &=\xt_i \otimes \Pi_\mathcal{S'} \otimes \id_f +  \tilde{Z}_i \otimes \bar{\Pi}_\mathcal{S'} \otimes \id_f \, .
\end{align}
where $\tilde{Z}_i$ commutes with $Z_i$. The commutator is then
\begin{align}
    [U^\dagger \xt_f U ,Z_i]&=[\xt_i \otimes \Pi_\mathcal{S'} \otimes \id_f +  \tilde{Z}_i \otimes \bar{\Pi}_\mathcal{S'} \otimes \id_f  ,Z_i]\\
    &=[\xt_i \otimes \Pi_\mathcal{S'} \otimes \id_f ,Z_i]\\
    &=2\yt_i \otimes \Pi_\mathcal{S'} \otimes \id_f, 
\end{align}
so $\norm{[U^\dagger \xt_f U ,Z_i]}_p $ equals $2\cdot D^{(-L+2)/p} \abs{\mathcal{S'}}^{1/p}$ for even $D$ and $2\cdot (D-1) D^{(-L+1)/p} \abs{\mathcal{S'}}^{1/p}$, saturating the bound above with $\abs{\mathcal{S}}= D \abs{\mathcal{S}'}$.

\section{Non-Unitary Protocols}\label{sec:measurements}
More general than unitary state transfer protocols are those with mid-circuit measurements, or indeed those defined by any completely-positive trace-preserving map. Our results extend to such maps, as we demonstrate in this section, restricting to qubits for simplicity. By the Stinespring dilation theorem \cite{Stinespring_1955}, we can represent any quantum channel via a unitary on a larger space. To avoid overloading the term `ancilla', we refer to this additional Hilbert space as the extra register and denote it via $\mathcal{H}_\text{reg}$. Let any state transfer protocol $\Lambda$ be defined by 
\begin{equation}
    \Lambda(\rho) = \Tr_\text{reg}[ W (\rho \otimes \dyad{\perp}_\text{reg}) W^\dagger],
\end{equation}
where $\dyad{\perp}_\text{reg}$ is the arbitrary pure initial state of the workspace. 
The state transfer conditions Eqs.~(3) and (4) in the main text still apply, so a $\mathcal{S}$-robust state transfer protocol must achieve 
\begin{align}
W^\dagger X_f W &=X_i S_x \quad \text{and} \\
W^\dagger Z_f W &=Z_i S_z,
\end{align}
where $S_{x,z}$ are stabilizer unitaries on the larger Hilbert space: 
\begin{equation}\label{dilatedstabilizers}
    S_x (\ket{\psi}_i \otimes \ket{\Phi}_{\rm anc} \otimes \ket{\perp}_{\rm reg})  = S_z (\ket{\psi}_i \otimes \ket{\Phi}_{\rm anc} \otimes \ket{\perp}_{\rm reg}) = \ket{\psi}_i \otimes \ket{\Phi}_{\rm anc} \otimes \ket{\perp}_{\rm reg} \quad \forall \ket{\Phi}\in \mathcal{S}, \forall \ket{\psi}.
\end{equation}
As a result, the derivation of Thm.~1 carries over to channels, with one difference: in our convention of renormalized $p$-norms, the factor due to the total Hilbert space dimension changes.
\begin{thm}\label{thmmainpnorm_channel}
    Any (possibly non-unitary) $\mathcal{S}$-robust state transfer protocol that uses a workspace $\mathcal{H}_\text{reg}$ satisfies
    \begin{equation}\label{cptpthm}
        \norm{[X_i S_x, Z_i]}_p \geq 2\cdot 2^{(-L+1)/p} \abs{\mathcal{S}}^{1/p} \abs{\mathcal{H}_\text{reg}}^{-1/p}.
    \end{equation}
\end{thm}

\subsection{Mid-Circuit Measurements}
So far we have assumed the state transfer protocols themselves are implemented perfectly and the only source of error is imperfect initialization of the ancillas. We can extend this assumption to also allow for certain kinds of errors in the implementation of the protocols. In particular, suppose we have a state transfer protocol utilizing mid-circuit measurements; such protocols are a common choice for enabling speedup when classical communication is effectively instantaneous~\cite{Briegel_1998}. 

Consider any one measurement in the circuit on a group of qubits. By Naimark's theorem \cite{Watrous2018}, the measurement can be realized as a projective measurement on a possibly larger set of qubits, which we'll refer to as the \emph{system}. Let projectors of this measurement be $\{P_i\}_{i=0}^{m-1}$. To read out the measurement result, we use a \emph{probe} of $\log_2 m$ qubits. Let $\ket{0}, ..., \ket{m-1}$ be the computational basis for the probe. We can define a unitary coupling the probe to the system such that reading out the probe in the computational basis effects the desired projective measurement. Assume the probe is initialized in $\ket{0}$. Define the unitaries $V_i=\dyad{i}{0}+\dyad{0}{i}+\sum_{j\neq0,i} \dyad{j}$. Then, the projective measurement is performed by applying the unitary 
\begin{equation}
    V = \sum_i P_i \otimes V_i
\end{equation}
and measuring the probe in the computational basis. 

Suppose there is a measurement error where the classical outcome gets forgotten: the system state is dephased, $\sum_i P_i \rho P_i$ and outcome $i$ is obtained with probability $1/m$. This is equivalent to the measurement register being initialized in the maximally mixed state.
Thus, if the state transfer protocol works despite this error, the stabilizers in Eq.~\eqref{dilatedstabilizers} must be stabilizers for all $m$ states of the measurement register, not just $\ket{0}$. This corresponds to increasing the norm in Eq.~\eqref{cptpthm} by a factor of $m^{1/p}$. 

\begin{thm}\label{thmmainpnorm_msmts}
    Consider a (possibly non-unitary) $\mathcal{S}$-robust state transfer protocol consisting of measurements and channels. Suppose the protocol is tolerant to errors in $n$ of the measurements, and that these measurements have $m_1,m_2,...,m_n$ outcomes respectively. As above, let the total workspace be $\mathcal{H}_\text{reg}$ such that the protocol can be represented as a Stinespring unitary on the lattice and workspace. Such a protocol must satisfy
    \begin{equation}
        \norm{[X_i S_x, Z_i]}_p \geq 2\cdot 2^{(-L+1)/p} \abs{\mathcal{S}}^{1/p}  \abs{\mathcal{H}_\text{reg}}^{-1/p} \prod_{i=1}^n m_i^{1/p}.
    \end{equation}
\end{thm}
In this sense, robustness to measurement errors in state transfer protocols can be treated on the same footing as errors in preparation of ancilla qubits.

\bibliography{mainbib}